\documentclass{article}
\usepackage[T1]{fontenc} 
\usepackage[utf8]{inputenc} 
\usepackage{ismir,amsmath,cite,url}
\usepackage{graphicx}
\usepackage{tabularray}

\usepackage{color}

\usepackage{tikz}
\usetikzlibrary{arrows.meta}

\usepackage{caption} 
\usepackage{microtype}
\usepackage[firstpage]{draftwatermark}
\definecolor{lightgray}{rgb}{0.9,0.9,0.9}
\definecolor{darkgray}{rgb}{0.4,0.4,0.4}
\SetWatermarkFontSize{12pt}
\SetWatermarkScale{1.1}
\SetWatermarkAngle{90}
\SetWatermarkHorCenter{202mm}
\SetWatermarkVerCenter{170mm}
\SetWatermarkColor{darkgray}
\SetWatermarkText{Late-Breaking / Demo Session Extended Abstract, ISMIR 2024 Conference}

\def\lbd{}	

\title{Language Models for Music Medicine Generation}

\multauthor
{Emmanouil Nikolakakis$^{*1}\thanks{* Equal contribution among authors.}$  \hspace{1cm} Joann Ching$^{*2}$ \hspace{1cm} Emmanouil Karystinaios$^{*2}$ } { \bfseries{Gabrielle Sipin$^3$ \hspace{1cm} Gerhard Widmer$^2$ \hspace{1cm} Razvan Marinescu$^1$}\\
$^1$ Department of Computer Science Engineering, University of California Santa Cruz, USA\\
$^2$  Institute of Computational Perception, Johannes Kepler University Linz, Austria\\
$^3$  Liberty Healthcare Corporation, San Francisco, USA\\
{\tt\small firstname.lastname@jku.at}
}

\def\authorname{E. Nikolakakis, E. Karystinaios, J. Ching, G. Sipin, G. Widmer, R. Marinescu}

\begin{document}

\maketitle
\begin{abstract}

Music therapy has been shown in recent years to provide multiple health benefits related to emotional wellness. In turn, maintaining a healthy emotional state has proven to be effective for patients undergoing treatment, such as Parkinson's patients or patients suffering from stress and anxiety. We propose fine-tuning MusicGen, a music-generating transformer model, to create short musical clips that assist patients in transitioning from negative to desired emotional states. Using low-rank decomposition fine-tuning on the MTG-Jamendo Dataset with emotion tags, we generate 30-second clips that adhere to the iso principle, guiding patients through intermediate states in the valence-arousal circumplex. The generated music is evaluated using a music emotion recognition model to ensure alignment with intended emotions. By concatenating these clips, we produce a 15-minute "music medicine" resembling a music therapy session. Our approach is the first model to leverage Language Models to generate music medicine. Ultimately, the output is intended to be used as a temporary relief between music therapy sessions with a board-certified therapist.


\end{abstract}
\section{Introduction}\label{sec:introduction}

Music therapy has proved to be a highly effective alternative treatment to traditional medication since the late nineteenth century. For most of the twentieth century, music therapy's positive effects in treating patients were investigated empirically, and the results were presented from a qualitative perspective. Nevertheless, recent advances in sensor technology have allowed clinicians to perform experiments that measure changes in heart rate, electromyogram, respiration, and skin conductance. Moreover, neuroimaging methodologies, such as Positron Emission Tomography and functional Magnetic Resonance Imaging \cite{blood2001intensely} enable the visualization of the signals transmitted between neurons and the cerebral blood flow within the brain during sessions. Therefore, the patient's response to music has been studied from a cognitive perspective.

With the advent of generative models such as Transformers and Diffusion models, various works have investigated the quality and diversity of AI-generated music for different applications \cite{huang2018musictransformer, wang2024WholeSongHG, jeong2018VirtuosoNet}. Some examples of tasks that can be accomplished involve style change~\cite{wu2023musemorphose}, orchestration~\cite{zhao2023q}, and novel music generation pieces from a given text prompt~\cite{copet2024simple}.

Following the popularity of generative AI, Williams \emph{et al.} \cite{williams2020use} evaluated the effectiveness of a Markov model to generate therapeutic music for a patient by measuring their Galvanic Skin Response at any given moment and confirmed its effects. Following those results, Hou \emph{et al.} \cite{hou2022ai} and Li \emph{et al.} \cite{li2022long} presented two long-short term memory (LSTM) based generative models adapted for music therapy given a treatment scenario. More recently, music generation in the symbolic domain has shown promising results for music therapy \cite{qiu2023generated}.

In this work, we propose a generative music medicine model\footnote{Demo: \url{https://tinyurl.com/gen-musmed}} by fine-tuning MusicGen~\cite{copet2024simple} using low-rank decomposition~\cite{hu2021lora} with prompts that contain emotion labels. 
We aim to generate a ``therapy session'' that follows the \textit{iso principle}~\cite{davis2008introduction}, which aims to alter a person's mood by playing music matching their current mood and then gradually shifting to music that represents a desired positive state. 
Iso is a commonly used practice in music therapy, with its effectiveness demonstrated in recent studies~\cite{heiderscheit2015use, starcke2021emotion}. 
We use the term \textit{music medicine} for the output of our model, as music therapy is always conducted in coordination with a licensed music therapist~\cite{stegemann2019music}. 

\begin{table*}[htbp]
\centering
\begin{tabular}{l | c c c |c c}
  &  CLAP $\uparrow$ & AUPRC $\uparrow$ & Ham Score $\uparrow$ & Time (min.) $\downarrow$ & Parameters $\downarrow$ \\ 
 \hline
 MusicGen-Large & 36.8 (+9.85\%) & 0.144 (+19.37\%) & 0.957 (+1.92\%) & $100 \pm 5$ & $3.3*10^{9}$ \\ 
 \hline
 MusicGen-Medium  & 38.7 (+2.97\%) & 0.06 (+6.00\%) & 0.929 (+0.96\%) & $88 \pm 10$ & $1.5*10^{9}$ \\
 \hline
 MusicGen-Small  &  33.1 (+19.12\%) & 0.031 (-3.44\%) & 0.942 (+0.95\%) & $48 \pm 2$ & $0.3*10^{9}$ \\
 \hline
\end{tabular}
\caption{
Evaluation scores for the three fine-tuned models and their percentage improvements over the non-finetuned version. The time column shows the average inference time for a 15-minute music session.}
\label{tab:baselines}
\end{table*}

\section{Methodology}\label{sec:page_size}

\subsection{Fine-tuning MusicGen}
We leverage the MusicGen model~\cite{copet2024simple}, a single Language Model (LM) for continuous conditional music generation. MusicGen was trained on 20K hours of licensed instrument-only music and approximately 400K tracks, thus creating an awareness of different genres and instruments. We fine-tune MusicGen with the MTG-Jamendo Dataset \cite{bogdanov2019mtg}, specifically using the subset where mood/theme tags are available, creating an emotion-aware music generation model. 

\begin{figure}
    \centering
    \begin{tikzpicture}[>=stealth, font=\small, scale=0.95]
    \draw (0,0) circle (2.5cm);
    \draw[->] (-2.8125,0) -- (2.8125,0) node[right, font=\footnotesize] {VALENCE};
    \draw[->] (0,-2.8125) -- (0,2.8125) node[above, font=\footnotesize] {AROUSAL};
    \node[font=\footnotesize] at (2.625,0.1875) {+};
    \node[font=\footnotesize] at (-2.625,0.1875) {-};
    \node[font=\footnotesize] at (0.1875,2.625) {+};
    \node[font=\footnotesize] at (0.1875,-2.625) {-};
    \node[font=\footnotesize] at (0.5,2.3) {alert};
    \node[font=\footnotesize] at (1.25,1.8125) {excited};
    \node[font=\footnotesize] at (1.75,1.0625) {elated};
    \node[font=\footnotesize] at (1.9,0.375) {happy};
    \node[font=\footnotesize] at (1.8,-0.1875) {contented};
    \node[font=\footnotesize] at (1.75,-1.0625) {serene};
    \node[font=\footnotesize] at (1.25,-1.6875) {relaxed};
    \node[font=\footnotesize] at (0.375,-2.2) {calm};
    \node[font=\footnotesize] at (-0.5,-2.1875) {bored};
    \node[font=\footnotesize] at (-1.3625,-1.25) {depressed};
    \node[font=\footnotesize] at (-2.125,-0.375) {sad};
    \node[font=\footnotesize] at (-2.075,0.1875) {upset};
    \node[font=\footnotesize] at (-1.55,1.0625) {stressed};
    \node[font=\footnotesize] at (-1.05,1.7125) {nervous};
    \node[font=\footnotesize] at (-0.4125,2.2125) {tense};
    \end{tikzpicture}    
    \caption{Emotion Circumplex Model~\cite{posner2005circumplex}}
    \label{fig:circumplex}
\end{figure}
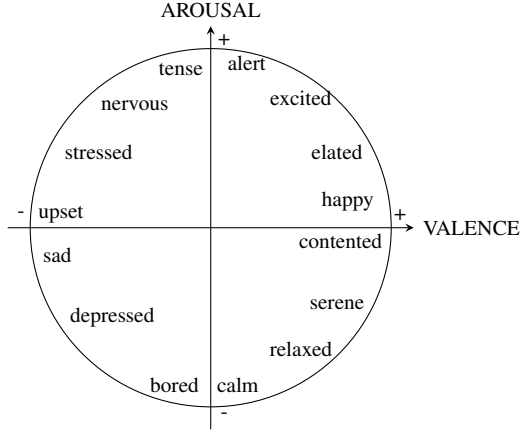

We use parameter-efficient fine-tuning training with Low-Rank Adaptation (LoRA)~\cite{hu2021lora} to fine-tune the MusicGen model without adapting the original weights provided by~\cite{copet2024simple}. As each piece in the subset is associated with mood tags, we can map them to specific emotions from the circumplex~\cite{posner2005circumplex} (Fig.~\ref{fig:circumplex}). A pile sorting experiment was conducted to assign an individual mapping of moods to emotions by three independent researchers and one by Anthropic's Claude AI. We obtained an inter-rater agreement score (Fleiss Kappa) of $0.2471$, confirming that our individual ranking corresponds to a fair agreement. Finally, the mapping is verified by a music therapist.  

We validate the fine-tuning between a set number of iterations by extracting the audio output's emotion using an emotion and theme recognition model~\cite{Mayerl2021mediaeval}. We then compare whether the extracted emotion matches the intended emotion of our prompt.


\subsection{Continuous Prompt Engineering}
Our model creates an audio output of around 15 minutes by concatenating multiple 30-second audio clips generated at each mood state. The audio clips are each generated with a given prompt 
to the fine-tuned model containing comma-separated tags concerning the mood tag, emotion, instrumentation, and genre (e.g. "sad, piano, classical"). To follow the iso principle, we determine a path between the initial and desired state, generating audio segments along the intermediate emotional states. Our approach is designed to begin with a text input for the first time step. For the following time steps, the model's input consists of part of the audio output of the previous time step, along with a text prompt consisting of the subsequent emotional state, the preferred instrument, and the genre of music. 
We introduce a temperature variable that increases the chance of changing the instrumentation and genre between audio clips according to the statistical distribution of these labels for a given mood in the MTG-Jamendo Dataset subset used for fine-tuning. 
To ensure high-quality generation, we trim any silence at beginning and end of the 30-sec generated clips before using them for conditioning the next generation.

After obtaining the generated clips, we normalize them individually. Subsequently, we crossfade them with an overlap of a quarter length from the previous one, therefore combining them into a single audio. We further post-process the concatenated audio by applying a high-pass filter and dynamic denoising using spectral gating. 
Therefore, we aim to have a final audio output that maximizes auditory pleasure and comfort for the individual undergoing the therapy.

\section{Results}\label{sec:typeset_text}

We fine-tuned and experimented with three different variations of MusicGen: 1) MusicGen-Small, 2) MusicGen-Medium, and 3) MusicGen-Large~\cite{copet2024simple}. After experimenting, we opted to fine-tune the model for 2 epochs with a learning rate of $7*10^6$ and a linear decay scheduler.
All training and inference were performed on a single NVIDIA A40 GPU.

The generated output from each model successfully traverses states in the circumplex, starting and ending at the initial and desired state, respectively. To quantitatively evaluate the output (see Table \ref{tab:baselines}), we use the following metrics: 1)  Contrastive Language-Audio Pretraining (CLAP)~\cite{elizalde2023clap}, which evaluates the relevance of the audio output to the text prompt input, 2) Area under the precision-recall curve (AUPRC) and 3) Hamming Score; the latter two are used to evaluate the mood of the generated audio segment by passing it to an emotion recognition model.

\section{Conclusions}

This work presented a novel approach to generative music medicine. 
The generated audio follows the iso principle, a well-established methodology in music therapy, dynamically adapting instruments and genres to reflect the patient's evolving emotional state. All steps of our work have been evaluated and approved by a licensed music therapist. Our future direction involves conducting a user study to evaluate the model's effectiveness in mental wellness improvement. Moreover, we wish to make a more controllable system by passing continuous coordinates on the circumplex as input to the prompt rather than discrete states to generate a smoother transition between states and audio segments.

\section{Acknowledgements}
Our implementation functions as a proof of concept, but it must still be evaluated experimentally before being clinically applied as medical treatment. This work is supported by the European Research Council (ERC) under the EU’s Horizon 2020 research \& innovation programme, grant agreement No. 101019375 (Whither Music?).

\bibliography{ISMIRtemplate}

%
%
%
%
%

\end{document}